\documentclass[review]{elsarticle}
\usepackage{amsmath}

\usepackage{arydshln}
\usepackage{lineno,hyperref}
\usepackage{algorithmic}
\usepackage{algorithm}
\usepackage{subfigure}
\modulolinenumbers[5]

\begin{document}

\begin{frontmatter}

\title{Agent-Based Modelling of Malaria Transmission Dynamics}
\tnotetext[mytitlenote]{Corresponding author: b.modu@bradford.ac.uk}

\author{Babagana Modu$^{1*}$, Nereida Polovina $^2$, Savas Konur$^1$}
\address{$^{*,1}$ Department of Computer Science, University of Bradford, United Kingdom}
\address{$^2$ Business School, Manchester Metropolitan University, United Kingdom}
%
%

\begin{abstract}
Recent statistics of malaria shows that over 200 million cases and estimated deaths of nearly half a million occur globally. Africa alone accounts for almost 90$\%$ of the cases. Several studies have been conducted to understand the disease transmission dynamics. In particular, mathematical methods have been frequently used to model and understand the disease dynamics and outbreak patterns. Although, mathematical methods have provided good results for homogeneous populations, these methods impose significant limitations for studying malaria dynamics in heterogeneous populations, a result of various factors, e.g. spatial and temporal fluctuations, social networks, human movements pattern etc. This paper proposes an agent-based modelling approach that permits modelling and analysing malaria dynamics for heterogenous populations. Our approach is illustrated using the climate and demographic data for the  Tripura, Limpopo and Benin cities. Our agent-based simulation has been validated against the reported cases of malaria collected in the cities mentioned. Furthermore, the efficiency of the proposed model has been compared with the mathematical model used as benchmark. A statistical test confirms the proposed model is robust and has potential for predicting the peak seasons of malaria. This potentially makes our methods a useful tool as an intervention mechanism, which will have impact on hospitals, healthcare providers, health organisations. 
\end{abstract}


\end{frontmatter}


\section{Introduction}
Despite continuous supports by the World Health Organisation (WHO) and other healthcare providers, malaria still remains as a global threat. Recent statistics on the spate of malaria shows that over 200 million cases and estimated deaths close to half a million occur every year. Africa alone accounts for almost 90$\%$ of the cases reported \cite{who2017malaria}. Beside poor healthcare facilities in Africa, the lack of efficient intervention mechanism for planning and managing disease outbreaks remains as a big challenge. 

Several studies (e.g., \cite{okuneye2017analysis, okuneye2018mathematical, nah2014malaria, nie2017roles,fan2010impact, agusto2015qualitative, paaijmans2009understanding, shapiro2017quantifying}) have been carried out using  (mathematical) compartmental modelling techniques to model and understand the disease transmission dynamics, which helped researchers  to investigate the progression of disease from a population perspective by characterising individuals according to their health status (e.g., susceptible, infected and recovered). Compartmental models are more often used compared to other types of models, such as stochastic models, complex network models, statistical process control models, spatial models and machine learning-based models \cite{siettos2013mathematical}. This is mainly because of their effectiveness in tracking disease progression and thereby transforming compartments into differential equations. Although compartmental models are powerful tools to investigate the disease dynamics, they impose a severe constraint, i.e. assuming homogeneity across the human population  \cite{herzog2017mathematical}. This means that each and every individual in the population has constant rates of infection, recovery and immunity loss. 

As a matter of fact, the human population is heterogeneous, and influenced by a broad spectrum of factors, including age, sex, spatial and temporal changes, human movement patterns, and social network patterns  \cite{kong2016modeling} amongst others. This limitation makes compartmental models inadequate for capturing the heterogeneities arising from the population dynamics of malaria. To address the limitations, we propose an agent-based modelling approach that alleviates the limitations imposed by compartmental models and permits modelling and analysing malaria dynamics for heterogenous populations.

Agent-based models (ABMs) are computational modelling tools consisting of agents, which communicate to each other within their environment and behave according to pre-defined rules. ABMs are powerful due to their stochasticity, spatial explicitness, and discrete-time-based simulation where each agent interacts in space and time \cite{hackl2019epidemic}. ABMs often work as a bottom-up modelling approach, as population-based behaviour emerges from interactions amongst autonomous agents \cite{willem2017lessons}. This characteristic means that ABMs are more flexible because of their ability to consolidate heterogeneous variables (e.g., host movement, heterogeneous implementation interventions) and stochastic (e.g., inter-patient variability at the time of infection, time to recovery, and the location of infection) \cite{smith2018agent}. Furthermore, ABMs allow a high degree of heterogeneity in the creation, disappearance and movement of a finite collection of discrete interacting individuals \cite{willem2017lessons}. The stochasticity of ABMs permits variation due to randomness, and thus more accurately mimics the transmission of malaria; it thereby reduces the effects of systematic preference amongst the agents \cite{smith2018agent}.

The agent-based malaria transmission models developed in this study enables us to investigate not only individual agents behaviour, but also how they communicate to each other according to predefined rules and their responses to climate factors. The agent-based models will be utilised to simulate the actual malaria cases in Tripura, Limpopo and Benin using the climate and demographic data obtained for these cities. Hence, the emerging results will be used to validate against the actual reported cases these cities. We also perform some statistical tests, such as \textit{t-test} for 2-independent samples and \textit{correlation} analysis, to evaluate the accuracy of the proposed model.
   
The paper is organised as follows: In Section~\ref{Models}, the malaria transmission framework together with agent-based and mathematical modelling approaches are discussed. Section~\ref{Methods} provides the vivid descriptions of the methodologies. Section~\ref{simul} presents the simulation analysis. Section~\ref{valid} presents and discusses our results. The conclusion and future work are summarised in Section~\ref{conclu}.   
   
\section{Model Formulation}\label{Models}
\begin{figure}[!ht]
\centering
\includegraphics[scale=0.5]{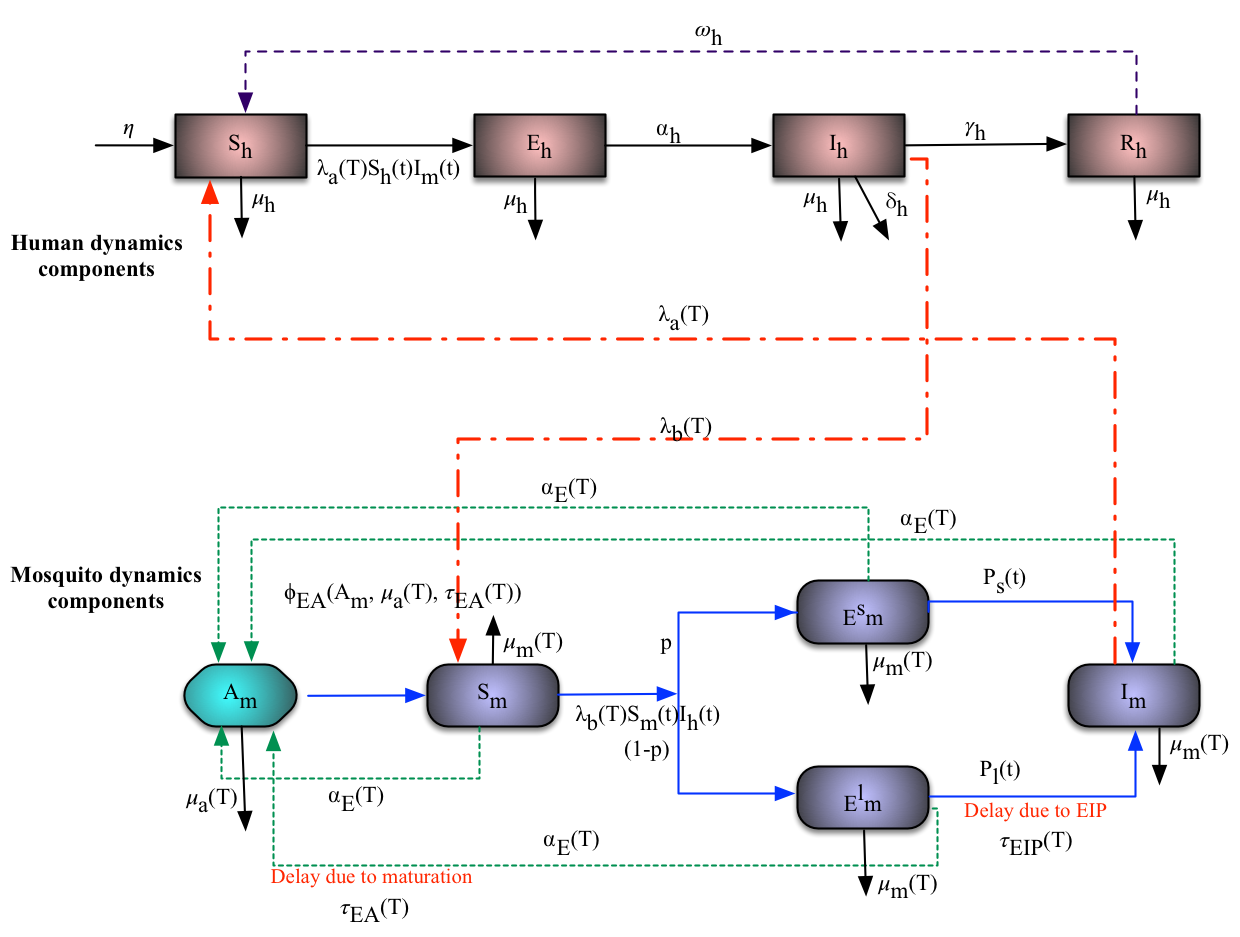}
\caption{Malaria transmission diagram.}
\label{tran-model}
\end{figure}

This section provides a theoretical background for compartmental modelling, agent-based modelling and their application to the study of malaria. We first present a malaria transmission model, as shown in Figure \ref{tran-model}, and describe its dynamics in human and mosquito populations; we also discuss the resultant complexity from the impact of temperature.

Malaria spreads into the human population through bites by the anopheles mosquito (female type), seeking human blood for nourishment and egg production. Figure \ref{tran-model} presents the compartments describing the human and mosquito dynamics, in which the human dynamics are structured using SEIR (susceptible, exposed, infected and recovered) attributes, and mosquito dynamics are structured using ASEI (aquatic, susceptible, exposed and infected) attributes. Temperature is incorporated into the model to account for its influence on the biting rate, survival rate, parasite development, juvenile maturation rate and mortality rate.

\begin{table}[!ht]
\caption{Model parameters and their definitions.}
\centering
\scalebox{0.75}{
\begin{tabular}{p{13cm}p{2cm}}
\hline
\textbf{Parameter} & \textbf{Symbol} \\ 
\hline
{Human recruitment rate} & $\eta$ \\ 
{Human immunity loss rate} & $\omega_h$ \\ 
{Human death rate} & $\mu_h$  \\ 
{Death rate of adult female mosquitoes} & $\mu_m$ \\
{Death rate of aquatic mosquitoes} & $\mu_a$ \\ 
{Egg deposition rate by adult female mosquitoes} & $\alpha_E$ \\
{Transition probability} & $p$ \\ 
{Transfer rate of exposed humans to infected class} & $\alpha_h$  \\ 
{Recovery rate for infectious humans to malaria infection} & $\gamma_h$\\ 
{Transfer rate of infected mosquitoes undergo latent period} & $\theta_m$ \\ 
{Delay due to malaria parasite incubation period} & $\tau_2$ \\ 
{Malaria transfer from infected mosquitoes to susceptible humans} & $a$ \\ 
{Malaria transfer from infected humans to susceptible mosquitoes} & $b$ \\ 
{Biting rate of mosquitoes on susceptible humans population} & $c_m$  \\ 
{Ratio of mosquitoes population to humans population} & $m$  \\ 
{Recruitment rate of adult female mosquitoes} & $\phi_{EA}$ \\ 
{Carrying capacity of immature mosquitoes} & $K_c$ \\
{Delay due to maturation rate of immature mosquitoes} & $\tau_1$ \\
{Malaria induced death rate of infectious humans} & $\delta_h$ \\
\hline
\label{definition of parameters}
\end{tabular}}
\end{table}


\subsection{Mathematical Model}\label{comp-model}
A compartmental model is a mathematical modelling technique that has long been used to investigate epidemics and public health policies. The `classical model' \cite{mandal2011mathematical} was the first mathematical technique that appearead in the literature that studied malaria transmission. Since then, several remarkable extensions have been made (e.g.,\cite{okuneye2017analysis, okuneye2018mathematical, nah2014malaria, nie2017roles,fan2010impact, agusto2015qualitative, paaijmans2009understanding, shapiro2017quantifying, macdonald1957epidemiology, aron1988mathematical, aron1982population, dietz1974malaria}) that build upon this model, addressing various emerging problems. In the mathematical modelling of malaria, a compartmental model (see Figure \ref{tran-model}) is used to describe the disease status transition using a set of differential equations \cite{hethcote2000mathematics}. The population within a particular compartment in a disease transmission model is assumed to be homogeneous, well-mixed and split (for instance, the SIR model) into compartments based on health status, e.g., susceptible, infectious and recovered. Hence, each compartment of the disease transmission model is defined by its own differential equations \cite{duan2015mathematical}. As shown in Figure \ref{tran-model}, the human population is split into four compartments depicted by the rectangular boxes labelled $S_h, E_h, I_h, R_h$ indicating susceptible, exposed, infectious and recovered humans, respectively. Similarly, the mosquito population is split into five compartments that include its juvenile stages, labelled $A_m, S_m, E^s_m, E^l_m, I_m$ indicating aquatic, susceptible, short-term exposed, long-term exposed and infectious, respectively. The differential equations describing the transition within the compartments in Figure \ref{tran-model} are skipped as outlining the mathematical models is not the focus of this paper. Table \ref{definition of parameters} detailed the definition of the parameters in Figure \ref{tran-model} while their values include their references (see Table \ref{values of parameters}). This model can be used to better understand the dynamics in a malaria transmission. However, mathematical models have proven to be suitable for modelling homogeneous systems at a macro-scale rather than at fine granular levels (with heterogeneity).

\subsection{Agent-Based Modelling}\label{agent-based}
In contrast to a mathematical model, an agent-based model is generally characterised by a bottom-up modelling approach. The agent-based approach enables individual agents to interact within their environment and behave according to predefined rules \cite{wang2013agent}. In addition, individual entities are represented by discrete autonomous agents communicating among themselves in a space to produce non-intuitive emergent patterns at the population level \cite{nejat2012agent}. Moreover, agent-based modelling represents purely rule-based algorithms, which start from scratch and continue until the desired model represents the real-world phenomena of interest. In general, agent-based modelling is characterised by its ability to capture heterogeneity, spatial and complex interactions, a micro-scale perspective, discrete time considerations and non-intuitiveness. All agents involved in the agent-based modelling of the malaria transmission dynamics in Figure \ref{tran-model} (including their detailed descriptions) are presented in the following subsections.

\subsubsection{Human Agents}\label{human-agent}
An infected human agent can transmit the malaria infection to susceptible mosquitoes provided that the incubation period of the malaria parasite in the human is complete. This period is called \textit{intrinsic incubation period} (IIP), which literally refers to the starting time when an infected mosquito has successfully infected a susceptible human, and the pathogen has started to greatly increase in number inside the host body \cite{jindal2017agent}. Subsequently, when the pathogen has multiplied and reached a certain threshold, malaria symptoms would then manifest. Furthermore, the chances of infection transmission would be very high during this period. The IIP is accounted for in Figure \ref{tran-model} and particularised as an exposed compartment in the human dynamics. In essence, agent-based modelling tries to realistically imitate the behaviour of individual agents and their interactions in the environment. The transmission of malaria is characterised by spatial and temporal considerations, in which the movement of people, whether short or long term, supports the spread of the malaria infection. For obvious reasons, people move from place to place within their environment for different purposes. In this paper, we consider the movement of people within their environment regardless of where they are heading, and thus observe the emerging patterns. Although mosquitos usually bite at night, other species bite during the day; however our concern in this paper is to track the pattern of the infection spread irrespective of day or night.

\subsubsection{Mosquito Agents}\label{mosquito-agent}
The mosquito agent is a carrier of the malaria parasite, and moves freely through a space in search of humans to bite, and thus transmits the parasite. Naturally, mosquitoes have certain characteristics that compose their life-cycle, which includes: biting rate, mortality rate, egg-deposition rate, birth rate and immature mortality rate. 

A malaria infection starts, if a susceptible mosquito bites an infected human, following which it becomes infected at a probability of $b$. Conversely, when such an infected mosquito bites a susceptible human, it infects the human with a probability of $a$. In Figure \ref{tran-model} we have illustrated the individual transition in the compartments (SEIR) and their respective probabilities.

\subsubsection{Pathogen Agents}\label{pathogen-agent}
The pathogen is a parasite causing the malaria infection, which in a biological sense is called the \textit{plasmodium species}. \textit{Plasmodium} transfers to the susceptible mosquito through biting an infected human and by sipping blood that contains the pathogen. Hence, the mosquito will become infected but not capable of transmitting the malaria infection to susceptible humans until the ingested blood containing the pathogen has developed. This will then follow some developmental stages before the mosquito becomes infectious. The time it takes for the pathogen to complete its development inside the mosquito is called the \textit{extrinsic incubation period} (EIP). This period is sensitive to environmental temperature \cite{mordecai2013optimal}, meaning at a relatively high or low temperature spectrum the EIP could be shorter or longer, respectively.

\subsubsection{Environment Agents}\label{environ-agent}
The environment as an agent leverages the spread of malaria and is ever-changing in space and time as the climate changes. The environment plays a vital role in the transmission of malaria, and enables the movement of people and mosquitoes, providing mosquitoes with breeding sites for egg deposition and the maturation of juvenile mosquitoes. In order to realistically represent the spatial movement of humans and mosquitoes, an artificial environment was created in the \textit{Netlogo} platform (see Figure \ref{setup_abm}), in which both humans and mosquitoes are displayed in the environment representing the spatial distribution of agents in a town, city, or community settlement.

\section{Materials and Methods}\label{Methods}
This section presents a vivid description of the methodologies used to achieve the aim of this study. Three cities were selected, each from a different country; these were Tripura district in India, Limpopo province in South Africa and Benin city in Nigeria. World map data \cite{gadm2019} were deployed to produce the maps of the three cities. However, since the entire city regions were difficult to handle or study, we resorted to the reduction of complexity in order to alleviate the computational task.

A statistical technique was adopted to reduce the computational task and guide the selection of the best performing model. The tools for simulation and computational functions of the temperature-dependent parameters of the malaria transmission model in Figure \ref{tran-model} are also presented.

\subsection{Case Study}\label{case-study}
This paper aims to investigate the dynamics of malaria in human and mosquito populations through agent-based modelling and mathematical modelling. The model developed will be validated against reported cases of malaria for different populations. To do this, we used data from three cities in different countries, including Tripura district in India, Limpopo province in South-Africa and Benin city in Nigeria. These countries are known for their malaria endemic status \cite{gething2011new}, and apparently have different seasons for transmission, climate patterns, parasites and vector species. For instance, Tripura district is known for its high malaria incidence, and its predominant species of malaria parasites are the \textit{plasmodium falciparum}. This species alone is accountable for about 90$\%$ of the cases reported, and \textit{plasmodium vivax} comprises the remaining 10$\%$ \cite{dev2015malaria} of cases in the district. In Limpopo province, malaria is still endemic, as incidence in the area is characterised by the low altitude and climate \cite{blumberg2007malaria}; moreover, it is connected to other regions in sharing boundaries with some parts of Zimbabwe and Mozambique where malaria incidence is similarly high. Furthermore, the malaria season in Limpopo coincides with its warm and rainy summer that starts in September and goes through to May of the following year \cite{craig1999climate}. According to the World Health Organization (WHO), Nigeria is rated among the highest malaria endemic countries across the globe \cite{dawaki2016nigeria}, and Benin city is the capital of a state called Edo located in the southern part of Nigeria. The city has a tropical climate, which is characterised by a longer rainy season over a 12-month period and the average annual temperature is 26.1$^\circ$C. This rainfall pattern (with an average 2025mm annual precipitation) is regarded as the most likely influential climate factor leading to the high incidence of malaria cases.

\subsection{Sources of Data}\label{data-source}
The reported cases of malaria for Tripura district, Limpopo province and Benin city were taken from published sources \cite{dev2015malaria, communicable2019diseases, adeyemo2013incidence} respectively. Since the occurrence of malaria is connected to climate factors, we need such data for the computation of temperature-dependent parameters. In this paper, we use the average monthly temperature data, as temperature is the large-scale driver of malaria transmission since it influences the mosquito survival, the parasite development, its biting rate and the aquatic development of the juvenile mosquito. The following is the record of the average monthly temperature, for Tripura district \cite{dev2015malaria}, Limpopo province and Benin city \cite{climate2019data}. The temperature data were collected for the same period of time, namely a year, in which the cases of malaria were reported (see Table \ref{cases_temp_data} for detailed information). 

\begin{table}[!ht]
\centering
\caption{Monthly cases of malaria and temperature distribution.}
\scalebox{0.60}{
\begin{tabular}{|*{7}{c|}}  
\hline
\multicolumn{1}{|c}{Months} & \multicolumn{6}{|c|}{Cities} \\ \hline 
& \multicolumn{2}{|c}{Tripura district, 2011} & \multicolumn{2}{|c}{Limpopo province, 2015} & \multicolumn{2}{|c|}{Benin city, 2011} \\ \hline 
& Reported cases & Mean temperature & Reported cases & Mean temperature & Reported cases & Mean temperature \\ \hline
January & 240 & 18 & 863 & 27.4 & 58 & 26.4   \\ 
February & 298 & 29.95 & 1843 & 26.6 & 110 & 27.2  \\ 
March & 552 & 27.5 & 1588 & 25 & 199 & 27.4  \\ 
April & 254 & 27.9 & 411 & 22.3 & 258 & 27.5  \\ 
May & 1398 & 29.4 & 85 & 18.2 & 534 & 27  \\ 
June & 1817 & 29.1 & 38 & 15 & 512 & 25.6  \\ 
July & 1833 & 29.05 & 25 & 15 & 396 & 24.5  \\ 
August & 1760 & 29.35 & 49 & 17.9 & 787 & 24.5  \\ 
September & 1181 & 29.3 & 123 & 21.5 & 1092 & 24.9  \\
October & 684 & 27.55 & 192 & 24.2 & 129 & 25.9  \\ 
November & 614 & 23.15 & 144 & 25.9 & 201 & 26.7  \\ 
December & 431 & 17.95 & 83 & 26.9 & 54 & 26  \\ \hline
\end{tabular}}
\label{cases_temp_data}
\end{table}

\subsection{Relationship Between Malaria Occurrence and Temperature}\label{relationship}
\begin{figure}[!ht]
    \centering
    \includegraphics[scale = 0.50]{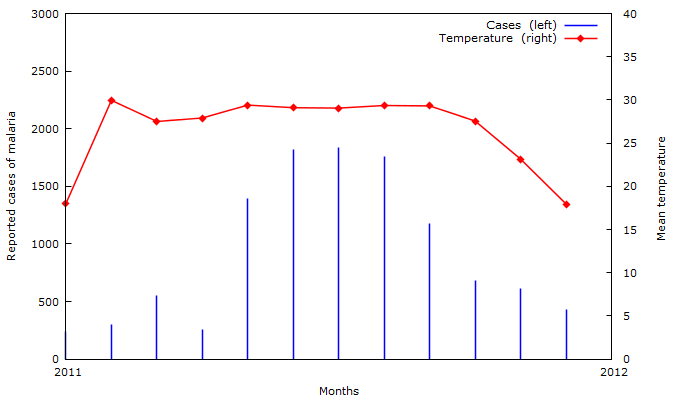}
    \caption{The plot describes the monthly cases of malaria and temperature distribution reported in Tripura district, India \cite{dev2015malaria}.}
    \label{cases_temp_plot-a}
\end{figure}

\begin{figure}[!ht]
    \centering
    \includegraphics[scale = 0.50]{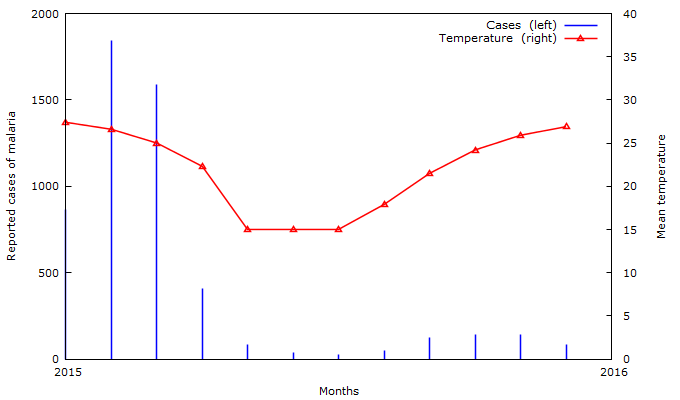} 
    \caption{The plot describes the monthly cases of malaria temperature distribution reported in Limpopo province, South Africa  \cite{climate2019data}.}
    \label{cases_temp_plot-b}
\end{figure}

\begin{figure}[!ht]
    \centering
    \includegraphics[scale = 0.50]{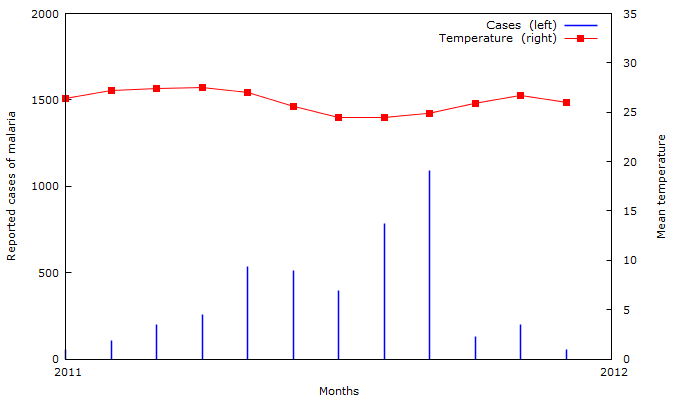} 
    \caption{The plot describes the monthly cases of malaria temperature distribution reported in Benin city, Nigeria \cite{climate2019data}.}
    \label{cases_temp_plot-c}
\end{figure}

As temperature is a large-scale driver of malaria transmission, it is crucial to check and substantiate the existing relationship. This would enable us to understand the spatial and temporal dynamics of malaria incidence. In Figures \ref{cases_temp_plot-a}--\ref{cases_temp_plot-c}, we present the relationships between the average monthly temperature distribution and the occurrences of malaria for Tripura district, Limpopo province and Benin city. As seen in Figure \ref{cases_temp_plot-a}, the relationship existing between the plots is relatively positive, showing that an increase in the mean monthly temperature causes a significant increase in the pattern of malaria incidence in Tripura. As shown in Figure \ref{cases_temp_plot-a}, the occurrences of malaria remain at high levels in the months starting from May until September. This shows a relatively unceasing pattern of malaria that spreads in Tripura, and could be possibly due to the average monthly temperature distribution that falls around the optimal temperature conditions (ranging between 25$^\circ$C--27$^\circ$C \cite{shapiro2017quantifying}) which favour \textit{plasmodium} parasite development. In Figure \ref{cases_temp_plot-b} the relationship is different from that observed in Figure \ref{cases_temp_plot-a}; this indicates the negative impact of a lower temperature regime due to the abundance of mosquitoes and parasite development. The parasite causing malaria will cease to develop when the temperature is below 14.5$^\circ$C for \textit{plasmodium vivax} and \textit{plasodium malariae}, and 16$^\circ$C for \textit{plasmodium falciparum} \cite{ohm2018rethinking}. Occurrences of malaria in Limpopo have lag effects of 1-2 months around the average monthly temperature distribution. This demonstrates the relationship between the malaria parasite incubation period and temperature \cite{nanvyat2018malaria}. The malaria season in Limpopo starts during the last quarter of a given year and continues until the first quarter of the following year. In Figure \ref{cases_temp_plot-c}, the pattern of malaria occurrence in Benin city is optimally driven by the average monthly temperature, which ranges from 25$^\circ$C--27$^\circ$C \cite{shapiro2017quantifying}. Moreover, the pattern of malaria incidence in Benin is perennial; its peak season starts February and lasts until November of the same year.

\subsection{Complexity Reduction}\label{reduction}
\begin{figure}[!ht]
    \centering
    \subfigure[]{\includegraphics[scale = 0.6]{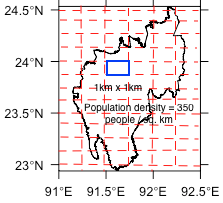}} 
    \subfigure[]{\includegraphics[scale = 0.6]{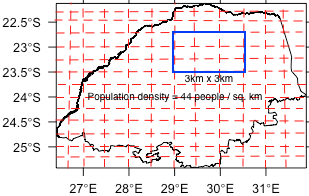}} 
    \subfigure[]{\includegraphics[scale = 0.6]{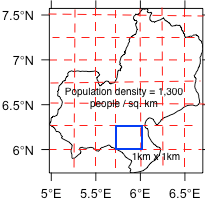}}
    \caption{Illustrates the scaled down area (a) Tripura, (b) Limpopo and (b) Benin.}
    \label{maps-trip}
\end{figure}

The entire regions covering the cities shown in Figures \ref{maps-trip}(a--c) are too large in terms of population size and land size. We have therefore scaled down the problem by defining ``unit zones'' in order to overcome the prevailing computational difficulties. Using the population size of the cities and land mass, the population density of the cities can be obtained from $\rho = N/A$, where $\rho$ = population density, $N$ = total human population and $A$ = total land area. Table \ref{table_population} presents the population density of the cities, showing that all cities have different densities. Hence, the cities were scaled down according to their densities, and the consideration was to study the areas covering the dimensions: 1km$\times$1km in Tripura, 3km$\times$3km in Limpopo and 1km$\times$1km in Benin (as indicated in Figures \ref{maps-trip}(a--c)). Based on the reduced dimensions of the cities, as shown in Figures \ref{maps-trip}(a--c), the sub-population, or target population, of the demarcated areas were as follows:~1:35 (representing 350 people), 1:4 (representing 400 people) and 1:13 (representing 1300 people) respectively.

%

\begin{table}[!ht]
\centering
\caption{Demographic information of the study areas.}
\scalebox{0.75}{
\begin{tabular}{|*{4}{c|}}  
\hline
\multicolumn{1}{|c}{} & \multicolumn{3}{|c|}{Cities} \\ \hline 
& \multicolumn{1}{|c}{Tripura district} & \multicolumn{1}{|c}{Limpopo province} & \multicolumn{1}{|c|}{Benin city} \\ 
\hline 
Population size & 3,673,917 & 5,554,657 & 1,495,800 \\
Land area & 10,492km$^2$ & 125,754Km$^2$ & 1,204Km$^2$ \\
Population density & 350/km$^2$ & 44/Km$^2$ & 1242/Km$^2$ \\
Target population & 350 & 400 & 1300 \\
Life Expectancy & 69 & 62.77 & 54.5  \\ 
\hline
\end{tabular}}
\label{table_population}
\end{table}

\subsection{Statistical tests}\label{stat-tests}
It is a verifiable fact that synthetic data are often generated to represent original data even though they are relatively different. However, George Box says \cite{box1976science} that \textit{all models are wrong, but some are useful}. Here, we focus on the application of agent-based modelling and mathematical modelling to study malaria dynamics in a population; we thus compare the results against the real cases. In fact, it is difficult to pinpoint the model that is best representation of the reported cases of malaria in each of the cities by merely looking at plotted patterns. To address this challenges, a \textit{t-test} for two independent samples \cite{kim2015t} and a \textit{product-moment} correlation \cite{modu2016data} were utilised. These statistical tests are robust for testing whether there is a significant difference between the means of the two independent samples (model generated cases and reported cases of malaria). For convenience, the mathematical representation of the \textit{t-test} is given by
\begin{equation}
t = \frac{\bar{X_a}-\bar{X_b}}{\sqrt{{\left(\frac{(N_a - 1)s^2_a +(N_b - 1)s^2_b}{N_a + N_b - 2}\right)}\left(\frac{1}{N_a} + \frac{1}{N_b}\right)}}
\end{equation}
where $\bar{X_a}$ and $\bar{X_b}$ are the means corresponding to sample $a$ and $b$ denoting simulated and real data respectively. Moreover, $s^2_a$ and $s^2_b$ are the sample variances, $N_a$ and $N_b$ are the sizes of the two independent samples, $N_a-1$, $N_b-1$ and $N_a + N_b-2$ are the degrees of freedom arising from individuals and combined samples respectively. Similarly, the population parameter $\rho_{x_jy}$, denotes a correlation coefficient between two variables, $x_j$ and $y$, and can be defined
\begin{equation}
\rho_{x_jy} = \frac{Cov(x_j, y)}{\sqrt{Var(x_j)Var(y)}},~\text{and}~j = 1,2,\cdots, m
\end{equation}
where $\rho_{x_jy} = \pm 1$ is constraint on the correlation coefficient, while $Cov(x_j,y)$, $Var(x_j)$ and $Var(y)$ are covariance of $x_j$ and $y$, variance of $x_j$ and variance of $y$ respectively. The statistical tests results are discussed in Section \ref{valid} while its summary is presented in Table \ref{inference_table}. 

\subsection{Parametrisation}\label{param}
Fundamentally, malaria transmission is leveraged by climate factors and, in particular, temperature is the large-scale driver of its transmission. As shown in Figure \ref{tran-model}, the mosquito related parameters, like the mosquito biting rate $c_m(T)$, adult mosquito mortality rate $\mu_m(T)$, immature mosquito mortality rate $\mu_a(T)$ and adult mosquito egg deposition rate $\alpha_E(T)$, all depend on temperature. Since there was no empirical evidence or values of these temperature-dependent parameters for the cities selected, the functional relationship between these parameters \cite{mordecai2013optimal} were utilised to determine the precise values corresponding to the demographic and climate information in Table \ref{table_population}. The temperature-dependent function of the parameters were described using the polynomial of degree two; their mathematical representations are shown in equation \ref{temp_para}. Similarly, the temperature-dependent linear function describing the immature mosquito mortality rate was defined in \cite{parham2012modeling}, and is shown in equation \ref{aquatic_death}. 

\begin{equation}
\begin{aligned}
c_m(T) &= -0.00014T^2 + 0.027T - 0.322\\
\mu_m(T) &= -\text{ln}(-0.000828T^2 + 0.0367T + 0.522)\\
\alpha_E(T) &= -0.153T^2 + 8.61T - 97.7\\
\end{aligned}
\label{temp_para}
\end{equation}
Similarly, the temperature-dependent linear function describing immature mosquito mortality rate is defined in~\cite{okuneye2017analysis}, given by 
\begin{equation}
\begin{aligned}
\mu_a(T) &= {1}/{8.560 + 2.654[1 + ({T}/{19.759})^{6.827}]^{-1}}
\end{aligned}
\label{aquatic_death}
\end{equation}

Using the temperature records of these cities (see \cite{dev2015malaria} for Tripura district and \cite{climate2019data} for Limpopo province and Benin city), we can obtain the values of the temperature-dependent parameters, which are presented in Table \ref{values of parameters}. The values are used in our simulations. Other parameters, e.g. the human birth or recruitment rate $\eta$, and the per capita human death rate $\mu_h$, can be calculated using the population size and human life expectancy of the cities (see Table \ref{table_population}). This information was accessed in the census database of the cities through online published sources (following the references \cite{census2019data,online2019tripura,online2019limpopo,online2019gauteng,online2019population,online2019benin}). The formulas used for computing the human birth and death rates are $\eta = \mu_h \times N$ and $\mu_h = 1/(LE \times I)$, where $N$ is the total human population size, $LE$ is the human life expectancy and $I$ is the index denoting the rate per month or day \cite{lou2010climate}.  

\begin{table}[!ht]
\caption{The parameter values and their ranges.}
\centering
\scalebox{0.80}{
\begin{tabular}{p{2cm}p{3cm}p{5cm}p{3cm}}
\hline
\textbf{Symbol} & \textbf{Baseline} & \textbf{Range} & \textbf{Reference} \\ 
\hline
 $\eta$ & $4\times10^{-5}$/day  & $(3.91-5)\times10^{-5}$/day  & \cite{nah2014malaria}\\ 
$\omega_h$ & $1.7\times10^{-5}$/day & $5.5\times10^{-5}-1.1\times10^{-2}$/day & \cite{agusto2015qualitative} \\ 
$\mu_h$ & $4\times10^{-5}$/day & $(3.42-3.91)\times10^{-5}$/day & \cite{agusto2015qualitative} \\ 
$\mu_m$ & $5\times10^{-2}$/day & $(4.76-7.14)\times10^{-2}$/day & \cite{agusto2015qualitative, niger2008mathematical}  \\
$\mu_a$ & $1.04\times10^{-1}$/day & $1\times10^{-3}-2\times10^{-1}$/day & \cite{okuneye2017analysis, agusto2015qualitative, lou2010climate} \\ 
$\alpha_E$ & $1.84$/day & $1-500$/day & \cite{okuneye2017analysis, agusto2015qualitative, lou2010climate}\\
$p$ & 0.25 & - & \cite{nah2014malaria}  \\ 
$\alpha_h$ & $5\times10^{-3}$ & $(2-7)\times10^{-3}$ & \cite{okuneye2017analysis} \\ 
$\gamma_h$ & $2.3\times10^{-3}$/day & $1.4\times10^{-3}-1.7\times10^{-2}$/day & \cite{agusto2015qualitative}\\ 
$\theta_m$ & $9.1\times10^{-2}$/day & $2.9\times10^{-2}-3.3\times10^{-1}$/day & \cite{agusto2015qualitative}  \\ 
$\tau_2$ & 10 & $10 - 14$/days & \cite{paaijmans2009understanding}  \\ 
$a$ & $2.4\times10^{-1}$/day & $7.2\times10^{-2}-6.4\times10^{-1}$/day & \cite{okuneye2017analysis, agusto2015qualitative,  niger2008mathematical, chitnis2006bifurcation} \\ 
$b$ & $2.2\times10^{-2}$/day & $2.7\times10{-3}-6.4\times10^{-1}$/day & \cite{ okuneye2017analysis, agusto2015qualitative, chitnis2006bifurcation} \\ 
$c_m$ & 0.29/day & $0.10 - 1.0$/day & \cite{okuneye2017analysis, agusto2015qualitative, niger2008mathematical, chitnis2006bifurcation}  \\ 
$m$ & 2 & - & \cite{nah2014malaria} \\ 
$\phi_{EA}$ & 0.343/day & $0.333 - 1.0$ & \cite{okuneye2017analysis} \\ 
$K_c$ & $4\times10^4$ & $50 - 3.3\times10^6$ & \cite{okuneye2017analysis, agusto2015qualitative} \\
$\tau_1$ & $12$ & $10 - 37$days & \cite{ okuneye2017analysis, mordecai2013optimal}  \\
$\delta_h$ & $3.454\times10^{-4}$ & $0 - 4.1\times10^{-4}$/day & \cite{agusto2015qualitative} \\
\hline
\end{tabular}}
\label{values of parameters}
\end{table}
\newpage
\subsection{Simulation Toolkits}\label{toolkits}
Two simulation platforms were used: \textit{VenSim} \cite{vensim2019software} is for mathematical modelling and \textit{NetLogo}  \cite{wilensky2015introduction} is for agent-based modelling. However, agent- based models can also be implemented through programming languages, such as \textit{C}, \textit{Java} and \textit{Python}. The \textit{C} programming language was used to develop \textit{Repast}, \textit{Soar} and \textit{Swarm} platforms. These platforms are primarily designed for social sciences, general learning problems and general purpose agent-based systems, respectively \cite{abar2017agent}; however, \textit{Java} is a versatile programming language used in building many platforms, including: \textit{AnyLogic}, \textit{Cougaar}, \textit{JADE}, \textit{MASON}, \textit{Repast}, \textit{SARL}, \textit{Soar}, \textit{Sugarscape} and \textit{Swarm}. In \textit{Python}, agent-based models are implemented in a framework called \textit{Mesa}, a modular framework for building, analysing and visualising agent-based models \cite{cardinot2019evoplex}.

\section{Experimentation}\label{simul}
In this section, we present our experiments by running the malaria model shown in Figure \ref{tran-model} using a mathematical modelling and agent-based modelling. A system dynamic modeller in the \textit{VenSim} platform was used to design the causal loop diagram shown in Figure \ref{sdm_plot}, and the simulation of the malaria transmission model in Figure \ref{tran-model}. The \textit{NetLogo} platform was subsequently utilised for agent-based modelling and simulation by tuning the values of the parameters in Table \ref{values of parameters}, e.g. average temperatures, population sizes, land masses, densities and human life expectancy. A detailed discussion of the experiments and processes is given below.

\subsection{VenSim Simulation}\label{ven-sim}
\begin{figure}[!ht]
\centering
\includegraphics[scale=0.48]{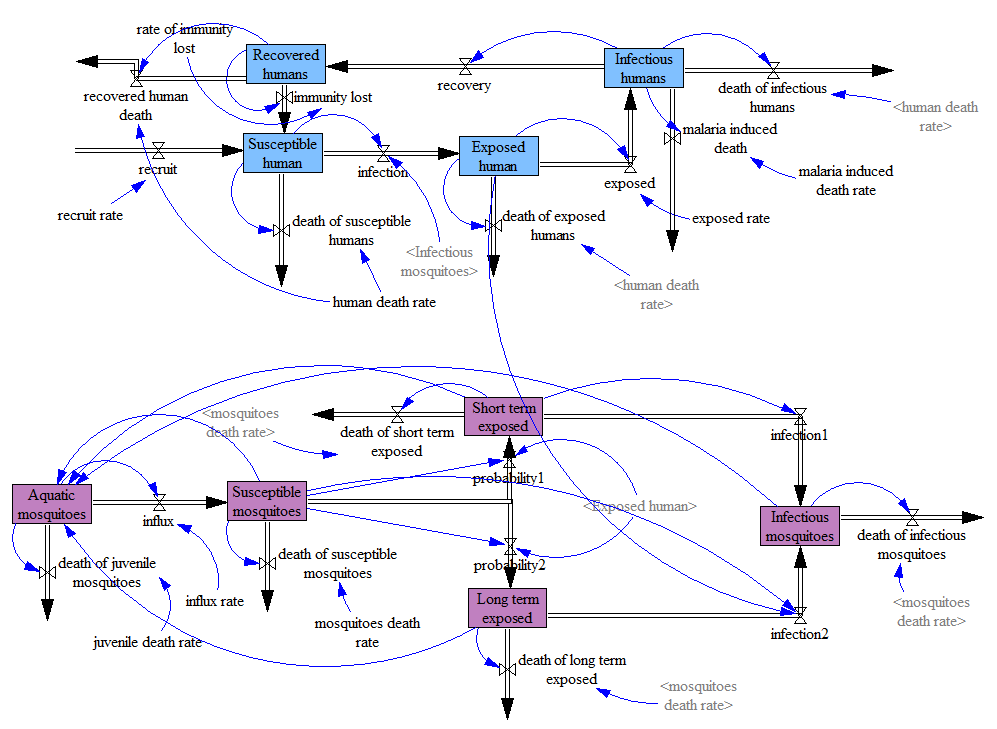}
\caption{Simulation of the model in Figure \ref{tran-model} using \textit{Vensim} system dynamic modeller.}
\label{sdm_plot}
\end{figure}

The digram in Figure \ref{sdm_plot} is a causal loop representation of the model in Figure \ref{tran-model} using the \textit{VenSim} system dynamic modeller. The values of the parameters in Table \ref{values of parameters} were picked and referenced to the cities' climate and demographic information and supplied within the system dynamic modeller's causal loop diagram in the \textit{VenSim} platform (see Figure \ref{sdm_plot}). Subsequently, the model was calibrated and the results were simulated within the ranges of the parameters; thus the dynamics of malaria transmission were generated for each of the cities. The results obtained for each of the cities are presented in Table \ref{simulation_table}, and further depicted in Figure \ref{simulation_results}. 

A numerical solution of mathematical modelling is a deterministic or non-probabilistic outcome in which, within a particular set of parameters, the results of the simulation remain consistent for any number of trials. However, in agent-based modelling, every trial of a simulation from the same set of parameters can lead to considerably different outcome.

\newpage
\subsection{NetLogo Simulation}\label{net-sim}
The \textit{NetLogo} platform recognises all mobile agents, like humans and mosquitos, as \textit{turtle}, and static agent, like environment, as  \textit{patches} \cite{wilensky2015introduction}. The visualisation dashboard provided in this platform is not sufficient to illustrate the number of human and mosquito agents. For this reason, the rescaled agent populations within the demarcated areas of the cities, as indicated in Figures \ref{maps-trip}(a--c), will now be used for this paper. Similarly, the mosquito population used for the simulation is actually difficult to determine as they are uncountable and are difficult to control. However, 1:2 ratio is often used for the human to mosquito population (as suggested by \cite{nah2014malaria,jindal2017agent}). For this reason, 1:40 mosquitoes (one real shaped mosquito in the \textit{NetLogo} environment represents 40 virtual mosquitoes) were represented in the \textit{NetLogo} space. Figure \ref{setup_abm} presents the initial setup of the agent-base model simulation interface of the malaria transmission model in Figure \ref{tran-model}. In order to realistically mimic human and mosquito disposition in the world, we spatially distributed all agents in the \textit{NetLogo} environment, while the sliders and plotting spaces were used for the parameters and outputs of the simulation, respectively.

\subsubsection{Creating Environment}\label{creat-envi}
\begin{figure}[!ht]
\centering
\includegraphics[scale=0.38]{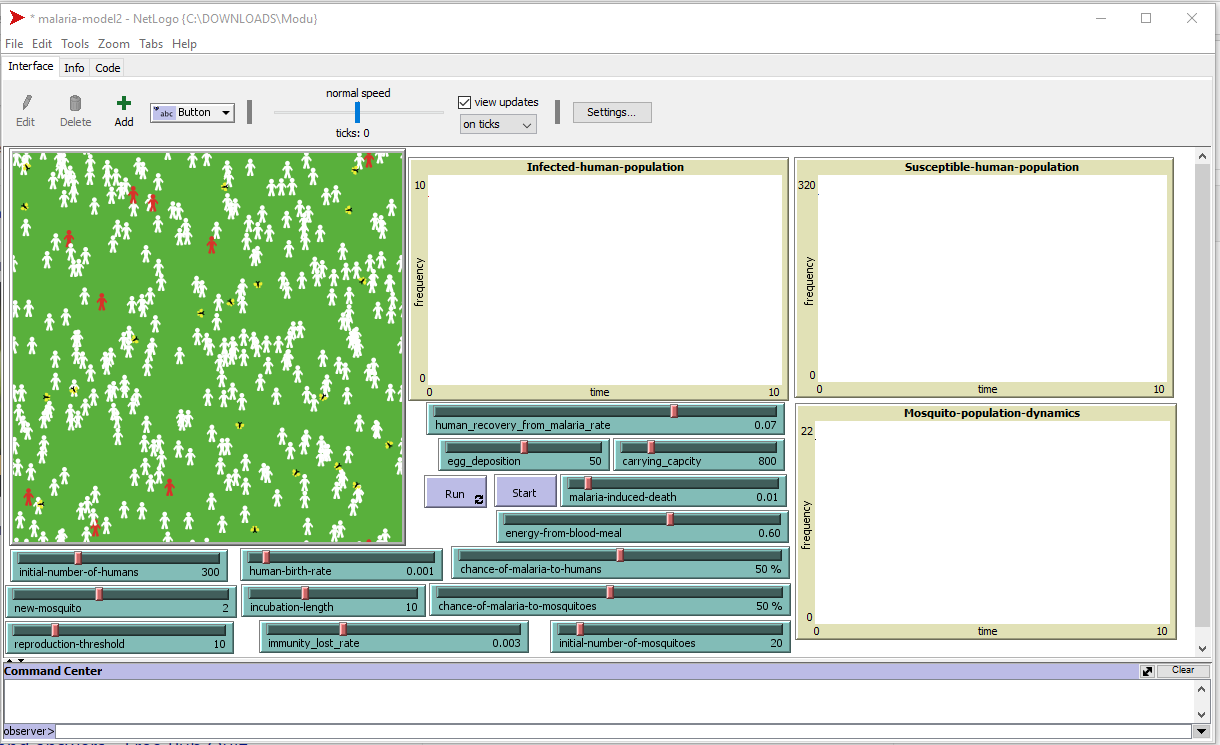}
\caption{A screenshot showing the interface of the ABM using the \textit{NetLogo} platform.}
\label{setup_abm}
\end{figure}

The setup of the agent-based model of malaria transmission in Figure  \ref{tran-model} using the \textit{NetLogo} platform is presented in Figure \ref{setup_abm}. We created the \textit{NetLogo} environment (\texttt{ask patches [set pcolor green]}), human agent and mosquito agent. The human and mosquito agents are depicted by \textit{person} and \textit{butterfly} shapes, respectively. Subsequently, the human and mosquito agents were initialised in accordance with their spatial dispositions in the environment by invoking [\texttt{random-xcor,random-ycor}], to account for systematic preferences during the simulation. In order to start malaria transmission within the created small environment, we assumed that few of the human agents have the malaria infection in their blood, and moreover, that humans are hosts of the pathogens causing the malaria \cite{pollitt2015existing}, called \textit{Plasmodium species}. Hence, for the malaria infection to effectively spread in the environment, we made the following assumptions:

\begin{itemize}
\item[i.] 10$\%$ of the human agents have the malaria infection.
\item[ii.] All mosquitoes are adults and susceptible at the initial time.  With the assumption that 10$\%$ of the human agents have malaria infection, we used a red colour in order to distinguish infected human agents and to avoid bewildering the susceptible counterparts.
\end{itemize}
Then, we invoked the following command to ask that 10$\%$ of the human agents become infected and are assigned a red colour, \texttt{if random-float initial-\\number-of-humans < 10}. Meanwhile, the remaining human agents were healthier and maintained their white colour, as shown in Figure \ref{setup_abm}. Since all agents were displaced randomly in the created \textit{NetLogo} environment, the next step was to define the sets of rules or instructions for the agents to follow and behave accordingly.

\subsubsection{Agents Procedure}\label{agent-proce}
The spread of malaria infection depends on how well the agents involved interact. People move from one place to another for several reasons, for example school, work, business or tourism. Thus, this movement increases the chance of becoming infected. Therefore, the agents have to move within the \textit{NetLogo} environment for the malaria infection to take effect (see Figure \ref{simulation}(a)), and the procedure \texttt{right-turn random 360$^\circ$, left-turn random 360$^\circ$ and forward 1} accounts for agents movements. Subsequently, the instruction follows that, if any of the mosquitos come into contact with red coloured humans, as the interaction progresses the particular mosquito will then be changed to a blue colour (see Figure \ref{simulation}(a)) indicating infected. The infected mosquito will not be capable of transmitting malaria to the susceptible human until it completes its extrinsic incubation period (EIP). This period mostly lasts for about 7-14 days \cite{ohm2018rethinking}; sometimes the mosquito and parasite species could also be another factor affecting the EIP. When the EIP is completed, the infected mosquito will then change colour to black (see Figure \ref{simulation}(a)) indicating infectious, and thus becomes a potential candidate for the spread of malaria. The aftermath is the spread of the malaria infection to the humans coloured white upon successful contact and bites by infectious mosquitoes. Consequently, infectious mosquitoes will pick human blood for nourishment and egg production. Regardless of whether the mosquito is infected or infectious, both will deposit eggs around swamp areas or water bodies, and thus new mosquitoes will be recruited (\texttt{hatch new-mosquito [set colour yellow forward 1]}). 

To better understand the procedural stages involved in the simulation cycles of the malaria transmission model (Figure \ref{tran-model}) using an agent-based model (as shown in Figure \ref{simulation}(a). We have developed \textbf{Algorithm} \ref{algo1} and \textbf{Algorithm} \ref{algo2} for the human and mosquito procedures.

\begin{algorithm}
\centering
\caption{Summary of the human procedures}
\begin{algorithmic}[1]
\STATE \textbf{let} initial human population be $i$  
\STATE \textbf{select} 10\% \textbf{random-float} of $i$ and assign colour red
\IF  {10\% of $i$ with colour red} 
        \STATE \textbf{set} status infected  
\ELSE
        \IF {$i$ coloured white}
              \STATE {\textbf{set} status susceptible}
              \STATE \textbf{for}($i$ in 1:$n$)
              \STATE {set $i$ to move \textbf{random left} or \textbf{random right} at 360$^\circ$}
              \STATE \textbf{repeat step 3}, for each iteration
              \STATE \textbf{then step 8}  
        \ENDIF
\ENDIF      
\end{algorithmic}
\label{algo1}
\end{algorithm}

\begin{algorithm}
\centering
\caption{Summary of the mosquito procedures}
\begin{algorithmic}[1]
\STATE \textbf{let} initial mosquito population be $j$
\STATE all $j^{\prime s}$ are assumed susceptible
\STATE \textbf{for} ($j$ in 1:n) 
\STATE \textbf{set} $j$ to fly \textbf{random left} or \textbf{random right} at 360$^\circ$ seeking $i^{\prime s}$ blood
\IF {any ($j$ in 1:$m$) bite any ($i$ in 1:$n$) with colour red} 
        \STATE \textbf{set} status infected for $j$ 
        \STATE \textbf{set} colour blue
        \STATE \textbf{repeat step 5}
\ELSE
        \IF {status is susceptible}
                \STATE otherwise \textbf{step 12}
                \STATE \textbf{for} ($j$ in 1:$m$) with status infected undergo extrinsic incubation period
\ELSE
        \IF {\textbf{step 12} is completed}
                \STATE \textbf{set} status infectious
                \STATE \textbf{set} colour black
                \STATE \textbf{repeat step 12}
                \STATE \textbf{repeat step 3 to 4} through \textbf{step 14} until all $j^{\prime s}$ are infectious
                \STATE \textbf{let} $\lambda$ be the average lifespan for $j^{\prime s}$
\ELSE
        \IF {lifespan for each $j^{\prime s}$ in \textbf{step 18} exceed $\lambda$}
                \STATE \textbf{set} status die
                \STATE \textbf{repeat step 3}
\ELSE   \IF {all $j^{\prime s}$ execute \textbf{step 21 to 22}}
                \STATE stop
             \ENDIF
         \ENDIF
      \ENDIF
   \ENDIF
\ENDIF 
\end{algorithmic}
\label{algo2}
\end{algorithm}
\clearpage

The setup has three components, which are:
\begin{itemize}
\item[i.] The main environment, which is the world where all agents would be spatially distributed accordingly.
\item[ii.] Three plane frames are meant to plot the simulation outputs in real-time as the agents interact in the environment according to predefined rules.
\item[iii.] Then widgets, of two kinds namely the sliders and buttons, are used to interact with the agents through the calibration of model parameters and start/run. 
\end{itemize}

\subsubsection{Testing veracity of the ABM}\label{veracity}
\begin{figure}[!ht]
    \centering
    \subfigure[]{\includegraphics[scale = 0.6]{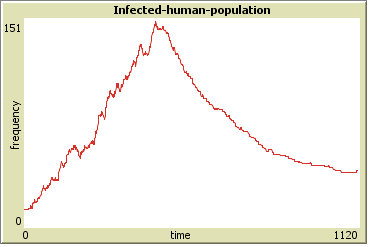}} 
    \subfigure[]{\includegraphics[scale = 0.6]{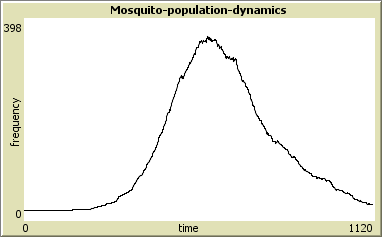}} 
    \subfigure[]{\includegraphics[scale = 0.6]{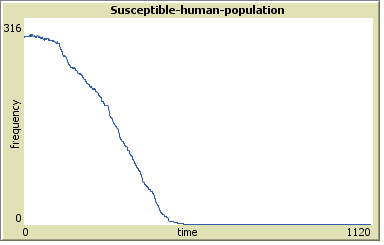}}
    \caption{This plots show (a) pattern of malaria’s infectiousness, (b) mosquito dynamics and (c) human susceptibility to malaria.}
    \label{simulation}
\end{figure}

To test the agent-based model shown in Figure \ref{setup_abm}, we performed a trial simulation and ran it 50 times by varying the referenced parameters in the sliders. We observe the pattern it produces through the plotting spaces while all agents communicate with each other in the environment according to the predefined rules. The plots shown in Figures \ref{simulation}(a--c) describes the pattern of infectiousness in humans, in mosquitoes, and susceptible human dynamics, respectively. The hypothetical values of the parameters were calibrated in the sliders, to test the veracity of the agent-based model to the malaria transmission model in Figure \ref{tran-model}. The pattern produced in Figures \ref{simulation}(a--c) clearly manifested the general characteristics of an epidemic model's behaviour \cite{bakare2015optimal}.

As the season of malaria transmission usually falls within a year depending on the length of the rainfall season, we pre-set the simulation to run for a year in order to spot the peak malaria season. In Figure \ref{simulation}(a) a uni-modal peak season is connected with a period of abundance for mosquitoes and Figure \ref{simulation}(b) corroborates this. The results produced in Figures \ref{simulation}(a and b) further proved that the incidence of malaria is largely leveraged by the availability of adult mosquitoes \cite{parham2012modeling}.

In a population with a relatively low fertility and mortality rate, the pattern of malaria susceptibility in humans will decrease as the infection increases. Thus, Figure \ref{simulation}(c) confirms that a susceptible human population shows a decreasing pattern as time increases. This shows the robustness of an agent-based model in its ability to study the characteristics of individual agents involved in the phenomena and mimics its real-world scenario.


%


By turning the values of the parameters within their ranges (Table \ref{values of parameters}) and with reference to the cities' climates and demographic information, the pattern of malaria transmission will then be resolved. For each of the cities we performed 100 runs of the simulation, and produced results on an average of the entire simulation cycle.

\section{Results Presentation and Validation}\label{valid}
\begin{figure}[!ht]
    \centering
    \subfigure[]{\includegraphics[scale = 0.37]{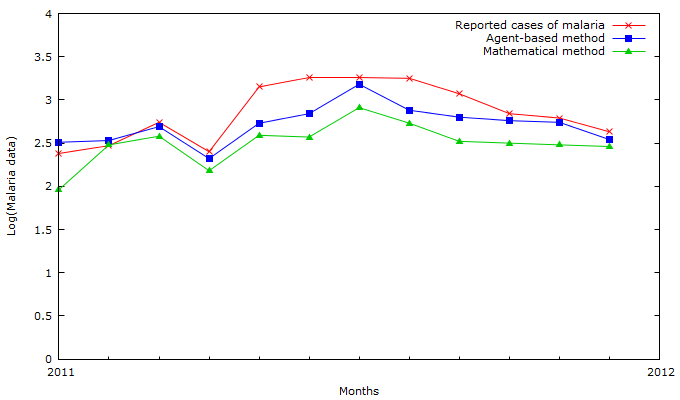}} 
    \subfigure[]{\includegraphics[scale = 0.37]{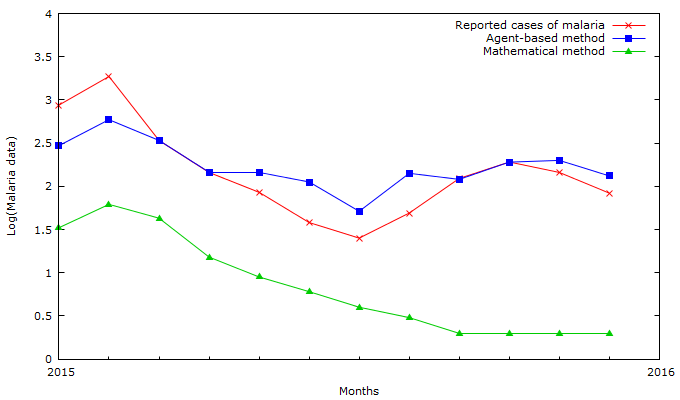}} 
    \subfigure[]{\includegraphics[scale = 0.37]{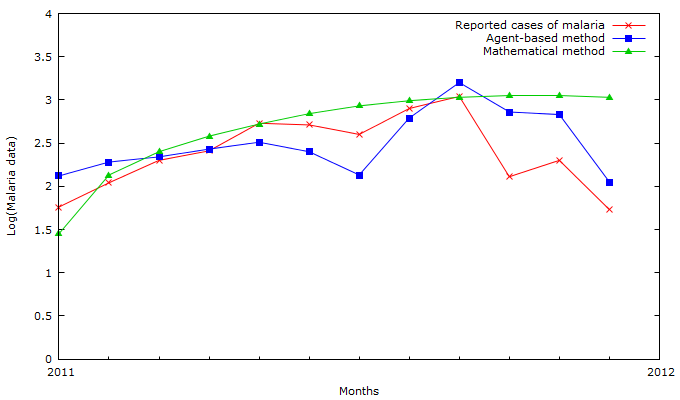}}
    \caption{The plots showing reported cases of malaria against model generated cases for (a) Tripura (b) Limpopo (c) Benin.}
\label{simulation_results}
\end{figure}

This section presents the results of the simulating model in Figure \ref{tran-model} through an agent-based and mathematical modelling approach. For each of the cities, we provided detailed results (see Table \ref{simulation_table}) and visualised them in Figure \ref{simulation_results}. The reported cases of malaria within the cities and the model's results are combined in Table \ref{simulation_table}. By looking at the patterns of the results compared with the cases, we noticed a wide variance, as indicated in the magnitude of the values. This made it difficult to critically comprehend the patterns. However, to make them consistent, since no zero instance was observed, a logarithmic transformation \cite{changyong2014log} technique was applied to normalise the results in Table \ref{simulation_table}. Subsequently, the log-transformed values in Table \ref{simulation_table} are plotted, as shown in Figure \ref{simulation_results}(a--c), and a detailed explanation follows.

Figure \ref{simulation_results}(a) presents the cases of malaria reported in the Tripura district against the model results produced using an agent-based model and mathematical model (see Table \ref{simulation_table}). The plots were produced using the two modelling approach, and indicate a certain strength of relationship within the occurrences of malaria in the district. The peak season of malaria incidence was predicted, as evidenced by the trough and node pattern produced in both the modelling approach, which is near-similar to the cases reported. The agent-based modelling and mathematical modelling performed well in predicting not only the pattern of malaria transmission in the district but also the season.

Similarly, Figure \ref{simulation_results}(b) shows a comparison between the cases of malaria reported in Limpopo province and the model generated results using an agent-based model and mathematical model (Table \ref{simulation_table}). The pattern produced by the plots in Figure \ref{simulation_results}(b) clearly shows that malaria occurrences in the province is represented well using the agent-based model results. However, the results produced using the mathematical model mimicked the pattern of the province malaria cases at the beginning of the simulation but later drifts down. This is because mathematical models are suitable for trend determinations and to investigate continuous phenomena. Hence, between the two modelling approaches, the agent-based model performed well in predicting the pattern of malaria incidence in the province.

Figure \ref{simulation_results}(c) presents the plots of the reported cases of malaria in Benin city together with model outcomes generated through an agent-based model and mathematical model (see Table \ref{simulation_table}). The plots in Figure \ref{simulation_results}(c) show a moderate relationship between the cases reported in Benin and the simulated results. In particular, the mathematical model produces a trend that follows a fluctuating pattern of malaria occurrences in the city. This shows that mathematical modelling is a good candidate for trend determination.
\begin{table}[!ht]
\centering
\caption{Reported cases of malaria of the three cities and the simulated results.}
\scalebox{0.51}{
\begin{tabular}{|*{19}{c|}}  
\hline
\multicolumn{1}{|c}{Months} & \multicolumn{18}{|c|}{Cities} \\ \hline 
& \multicolumn{6}{|c}{Tripura district, 2011} & \multicolumn{6}{|c}{Limpopo province, 2015} & \multicolumn{6}{|c|}{Benin city, 2011} \\ \hline 
& C & log(C) & A & log(A) & M & log(M) & C & log(C) & A & log(A) & M & log(M) & C & log(C) & A & log(A) & M & log(M) \\ \hline
Jan  & 240 & 2.38 & 320 & 2.51 & 91 & 1.96 & 863 & 2.94 & 293 & 2.47 & 33 & 1.52 & 58 & 1.76 & 133 & 2.12 & 28 & 1.45\\
Feb & 298 & 2.47 & 337 & 2.53 & 300 & 2.48 & 1843 & 3.27 & 594 & 2.77 & 61 & 1.79 &  110 & 2.04 & 191 & 2.28 & 134 & 2.13 \\
Mar & 552 & 2.74 & 493 & 2.69 & 384 & 2.58 & 1588 & 3.20 & 341 & 2.53 & 43 & 1.63 & 199 & 2.30 & 219 & 2.34 & 251 & 2.40 \\
Apr & 254 & 2.40 & 210 & 2.32 & 152 & 2.18 & 411 & 2.61 & 143 & 2.16 & 15 & 1.18 & 258 & 2.41 & 268 & 2.43 & 377 & 2.58 \\
May & 1398 & 3.15 & 538 & 2.73 & 388 & 2.59 & 85 & 1.93 & 143 & 2.16 & 09 & 0.95 & 534 & 2.73 & 324 & 2.51 & 523 & 2.72 \\
Jun & 1817 & 3.26 & 699 & 2.84 & 374 & 2.57 & 38 & 1.58 & 113 & 2.05 & 06 & 0.78 & 512 & 2.71 & 251 & 2.40 & 686 & 2.84 \\
Jul & 1833 & 3.26 & 1520 & 3.18 & 812 & 2.91 & 25 & 1.40 & 51 & 1.71 & 04 & 0.60 & 396 & 2.60 & 134 & 2.13 & 848 & 2.93 \\
Aug & 1760 & 3.25 & 750 & 2.88 & 543 & 2.73 & 49 & 1.69 & 142 & 2.15 & 03 & 0.48 & 787 & 2.90 & 610 & 2.79 & 986 & 2.99 \\
Sep & 1181 & 3.07 & 630 & 2.80 & 329 & 2.52 & 123 & 2.09 & 120 & 2.08 & 02 & 0.30 & 1092 & 3.04 & 1567 & 3.20 & 1078 & 3.03 \\
Oct & 684 & 2.84 & 575 & 2.76 & 315 & 2.50 & 192 & 2.28 & 192 & 2.28 & 02 & 0.30 & 129 & 2.11 & 720 & 2.86 & 1118 & 3.05 \\
Nov & 614 & 2.79 & 554 & 2.74 & 302 & 2.48 & 144 & 2.16 & 201 & 2.30 & 02 & 0.30 & 201 & 2.30 & 675 & 2.83 & 1111 & 3.05 \\
Dec & 431 & 2.63 & 347 & 2.54 & 289 & 2.46 & 83 & 1.92 & 132 & 2.12 & 02 & 0.30 & 54 & 1.73 & 112 & 2.05 & 1074 & 3.03 \\\hline
\end{tabular}}
\label{simulation_table}\\
\footnotesize{where: C = Reported cases of malaria, A = Agent-based method results, M = Mathematical method results and log(C), log(A), log(M) are their corresponding logarithmic transformations.}
\end{table}

%
\cleardoublepage
\subsection{Statistical Tests and Inferences}\label{inferences}
The plots in Figure \ref{simulation_results}(a--c) demonstrates a pattern that means it is difficult to recognise the best performing model in predicting the occurrence of malaria between the agent-based and mathematical models. A \textit{t-test} for two independent samples \cite{kim2015t} together with a \textit{correlation coefficient} \cite{modu2016data} were utilised and the best performing model was selected. The computational results of the parameters for the two statistical techniques were summarised and presented in Table \ref{inference_table}. 

\begin{table}[!ht]
\centering
\caption{Present the summarised results of the \textit{t-test} for two independent samples.}
\scalebox{0.75}{
\begin{tabular}{|*{7}{c|}}  
\hline
\multicolumn{1}{|c}{} & \multicolumn{6}{|c|}{Cities} \\ \hline 
& \multicolumn{2}{|c}{Tripura district} & \multicolumn{2}{|c}{Limpopo province} & \multicolumn{2}{|c|}{Benin city} \\ \hline 
t-statistic & C vs A & C vs M & C vs A & C vs M & C vs A & C vs M \\ \hline
t-value & 1.6378 & 2.9560 & 1.3444 & 2.3861 & 0.0850 & 3.4243  \\ 
p-value & 0.1157 & 0.0073 & 0.1925 & 0.0261 & 0.9330 & 0.0084  \\
r & 0.7732 & 0.7768 & 0.9784 & 0.9318 & 0.7176 & 0.5420 \\ 
Remark & Not sign & Sign & Not sign & Sign & Not sign & Sign  \\ \hline
\end{tabular}}
\label{inference_table}\\
\vspace{0.3cm}
\footnotesize{where: t-value = test-statistic, p-value = highest threshold probability for rejecting the null hypothesis, r = coefficient of correlation use to measure a degree of association, Not sign = indicating there is no significant difference and Sign = indicating there is significant difference.}
\end{table}

Based on the results presented in Table \ref{inference_table}, Figure \ref{simulation_results}(a) shows that there is  no significant difference on average between the reported cases of malaria and the results produced through an agent-based model ($p_{value} = 0.1157$, is greater than $\alpha = 0.005$). Similarly, it is also observed that there is a significant difference in the results produced through the mathematical model compared to the reported cases of malaria in Tripura ($p_{value} = 0.0073$, is less than $\alpha = 0.005$). A moderate degree of association was found (see Table \ref{inference_table}, $r = 0.7732$ and $r = 0.7768$) between the reported cases of malaria and the model-based results produced using the agent-based model and mathematical model, respectively. The results in Table \ref{inference_table} affirmed no significant difference between the reported cases of malaria and the results produced using the agent-based model in Limpopo and Benin ($p_{value} = 0.1925$, is greater than $\alpha = 0.005$ and $p_{value} = 0.9330$, is greater than $\alpha = 0.005$ respectively). However, the mathematical model results were found to be significantly different with the malaria cases reported in the two cities ($p_{value} = 0.0261$, is less than $\alpha = 0.005$ and $p_{value} = 0.0084$, is less than $\alpha = 0.005$ respectively). Furthermore, a strong correlation between the cases of malaria in Limpopo and the agent-based model cases was found ($r = 0.9784$ in Table \ref{inference_table}), and this result was also similar to those produced by the mathematical model ($r = 0.9318$ in Table \ref{inference_table}). However, in Benin the degree of association between the cases of malaria and the results produced using the agent-based model was moderate good compared to the mathematical model ($r = 0.7176$ and $r = 0.5420$ respectively).

In all the cities studied, the agent-based model performed well in predicting the occurrences of malaria incidence as opposed to mathematical model. However, the mathematical model was good for trend analysis and performed particularly very well for Tripura and Benin (see Figure \ref{simulation_results}(a and c)) compared to Limpopo (see Figure \ref{simulation_results}(c)). 

  
\section{Conclusion}\label{conclu} 
The unpredictability of malaria season driven by the factors influencing its transmission is a serious challenge. This paper addresses the limitations of mathematical modelling approach by deploying agent-based technique to the investigation of malaria transmission. The agent-based model was simulated and validated against the hospital reported cases, and also it precisions was compared against the mathematical model, used as benchmark. We observed that agent-based model proved to be robust in predicting season of malaria and the possible fluctuations. This results can help healthcare providers and policy makers with a schema showing how malaria invades population, so as to plan for feasible intervention. 

Our future work will be extending the agent-based model studied to investigate the insight into: (i) human movements pattern (ii) mosquito breeding sites.

\section*{Competing Interests}
The authors declare that they have no competing interests.
\section*{Author's Contributions}
Savas Konur and Nereida Polovina designed the experiments; Babagana Modu performed the experiments; Babagana Modu analysed the data; Babagana Modu, Savas Konur and Nereida Polovina analysed and evaluated the results; Babagana Modu wrote the paper; all authors edited the paper.
\section*{Acknowledgements}
B. Modu acknowledges the Tertiary Education Trust Fund (TetFund) Nigeria (in collaboration with Yobe State Univer- sity Damaturu) for sponsoring his PhD studies at University of Bradford. The work of S. Konur is supported by EPSRC (EP/R043787/1).

 

%

%
%


\bibliography{mybibfile}

\begin{thebibliography}{57}
\expandafter\ifx\csname natexlab\endcsname\relax\def\natexlab#1{#1}\fi
\providecommand{\url}[1]{\texttt{#1}}
\providecommand{\href}[2]{#2}
\providecommand{\path}[1]{#1}
\providecommand{\DOIprefix}{doi:}
\providecommand{\ArXivprefix}{arXiv:}
\providecommand{\URLprefix}{URL: }
\providecommand{\Pubmedprefix}{pmid:}
\providecommand{\doi}[1]{\href{http://dx.doi.org/#1}{\path{#1}}}
\providecommand{\Pubmed}[1]{\href{pmid:#1}{\path{#1}}}
\providecommand{\bibinfo}[2]{#2}
\ifx\xfnm\relax \def\xfnm[#1]{\unskip,\space#1}\fi
\bibitem[{who(2017)}]{who2017malaria}
\bibinfo{title}{World malaria report 2107: World health organization; 2017},
\newblock \bibinfo{journal}{\url{http://www.who.int/malaria/publications/
  world-malaria-report-2017/en/},~(accessed~10~March~2019)}
  (\bibinfo{year}{2017}).
\bibitem[{Okuneye and Gumel(2017)}]{okuneye2017analysis}
\bibinfo{author}{K.~Okuneye}, \bibinfo{author}{A.~B. Gumel},
\newblock \bibinfo{title}{Analysis of a temperature-and rainfall-dependent
  model for malaria transmission dynamics},
\newblock \bibinfo{journal}{Mathematical biosciences} \bibinfo{volume}{287}
  (\bibinfo{year}{2017}) \bibinfo{pages}{72--92}.
\bibitem[{Okuneye et~al.(2018)Okuneye, Abdelrazec, and
  Gumel}]{okuneye2018mathematical}
\bibinfo{author}{K.~Okuneye}, \bibinfo{author}{A.~Abdelrazec},
  \bibinfo{author}{A.~B. Gumel},
\newblock \bibinfo{title}{Mathematical analysis of a weather-driven model for
  the population ecology of mosquitoes},
\newblock \bibinfo{journal}{Mathematical Biosciences \& Engineering}
  \bibinfo{volume}{15} (\bibinfo{year}{2018}) \bibinfo{pages}{57--93}.
\bibitem[{Nah et~al.(2014)Nah, Nakata, and R{\"o}st}]{nah2014malaria}
\bibinfo{author}{K.~Nah}, \bibinfo{author}{Y.~Nakata},
  \bibinfo{author}{G.~R{\"o}st},
\newblock \bibinfo{title}{Malaria dynamics with long incubation period in
  hosts},
\newblock \bibinfo{journal}{Computers \& Mathematics with Applications}
  \bibinfo{volume}{68} (\bibinfo{year}{2014}) \bibinfo{pages}{915--930}.
\bibitem[{Nie and Xue(2017)}]{nie2017roles}
\bibinfo{author}{L.-F. Nie}, \bibinfo{author}{Y.-N. Xue},
\newblock \bibinfo{title}{The roles of maturation delay and vaccination on the
  spread of dengue virus and optimal control},
\newblock \bibinfo{journal}{Advances in Difference Equations}
  \bibinfo{volume}{2017} (\bibinfo{year}{2017}) \bibinfo{pages}{278}.
\bibitem[{Fan et~al.(2010)Fan, Liu, Van~den Driessche, Wu, and
  Zhu}]{fan2010impact}
\bibinfo{author}{G.~Fan}, \bibinfo{author}{J.~Liu}, \bibinfo{author}{P.~Van~den
  Driessche}, \bibinfo{author}{J.~Wu}, \bibinfo{author}{H.~Zhu},
\newblock \bibinfo{title}{The impact of maturation delay of mosquitoes on the
  transmission of west nile virus},
\newblock \bibinfo{journal}{Mathematical Biosciences} \bibinfo{volume}{228}
  (\bibinfo{year}{2010}) \bibinfo{pages}{119--126}.
\bibitem[{Agusto et~al.(2015)Agusto, Gumel, and Parham}]{agusto2015qualitative}
\bibinfo{author}{F.~Agusto}, \bibinfo{author}{A.~Gumel},
  \bibinfo{author}{P.~Parham},
\newblock \bibinfo{title}{Qualitative assessment of the role of temperature
  variations on malaria transmission dynamics},
\newblock \bibinfo{journal}{Journal of Biological Systems} \bibinfo{volume}{23}
  (\bibinfo{year}{2015}) \bibinfo{pages}{1550030}.
\bibitem[{Paaijmans et~al.(2009)Paaijmans, Read, and
  Thomas}]{paaijmans2009understanding}
\bibinfo{author}{K.~P. Paaijmans}, \bibinfo{author}{A.~F. Read},
  \bibinfo{author}{M.~B. Thomas},
\newblock \bibinfo{title}{Understanding the link between malaria risk and
  climate},
\newblock \bibinfo{journal}{Proceedings of the National Academy of Sciences}
  \bibinfo{volume}{106} (\bibinfo{year}{2009}) \bibinfo{pages}{13844--13849}.
\bibitem[{Shapiro et~al.(2017)Shapiro, Whitehead, and
  Thomas}]{shapiro2017quantifying}
\bibinfo{author}{L.~L. Shapiro}, \bibinfo{author}{S.~A. Whitehead},
  \bibinfo{author}{M.~B. Thomas},
\newblock \bibinfo{title}{Quantifying the effects of temperature on mosquito
  and parasite traits that determine the transmission potential of human
  malaria},
\newblock \bibinfo{journal}{PLoS biology} \bibinfo{volume}{15}
  (\bibinfo{year}{2017}) \bibinfo{pages}{e2003489}.
\bibitem[{Siettos and Russo(2013)}]{siettos2013mathematical}
\bibinfo{author}{C.~I. Siettos}, \bibinfo{author}{L.~Russo},
\newblock \bibinfo{title}{Mathematical modeling of infectious disease
  dynamics},
\newblock \bibinfo{journal}{Virulence} \bibinfo{volume}{4}
  (\bibinfo{year}{2013}) \bibinfo{pages}{295--306}.
\bibitem[{Herzog et~al.(2017)Herzog, Blaizot, and
  Hens}]{herzog2017mathematical}
\bibinfo{author}{S.~A. Herzog}, \bibinfo{author}{S.~Blaizot},
  \bibinfo{author}{N.~Hens},
\newblock \bibinfo{title}{Mathematical models used to inform study design or
  surveillance systems in infectious diseases: a systematic review},
\newblock \bibinfo{journal}{BMC infectious diseases} \bibinfo{volume}{17}
  (\bibinfo{year}{2017}) \bibinfo{pages}{775}.
\bibitem[{Kong et~al.(2016)Kong, Wang, Han, and Cao}]{kong2016modeling}
\bibinfo{author}{L.~Kong}, \bibinfo{author}{J.~Wang}, \bibinfo{author}{W.~Han},
  \bibinfo{author}{Z.~Cao},
\newblock \bibinfo{title}{Modeling heterogeneity in direct infectious disease
  transmission in a compartmental model},
\newblock \bibinfo{journal}{International journal of environmental research and
  public health} \bibinfo{volume}{13} (\bibinfo{year}{2016})
  \bibinfo{pages}{253}.
\bibitem[{Hackl and Dubernet(2019)}]{hackl2019epidemic}
\bibinfo{author}{J.~Hackl}, \bibinfo{author}{T.~Dubernet},
\newblock \bibinfo{title}{Epidemic spreading in urban areas using agent-based
  transportation models},
\newblock \bibinfo{journal}{Future Internet} \bibinfo{volume}{11}
  (\bibinfo{year}{2019}) \bibinfo{pages}{92}.
\bibitem[{Willem et~al.(2017)Willem, Verelst, Bilcke, Hens, and
  Beutels}]{willem2017lessons}
\bibinfo{author}{L.~Willem}, \bibinfo{author}{F.~Verelst},
  \bibinfo{author}{J.~Bilcke}, \bibinfo{author}{N.~Hens},
  \bibinfo{author}{P.~Beutels},
\newblock \bibinfo{title}{Lessons from a decade of individual-based models for
  infectious disease transmission: a systematic review (2006-2015)},
\newblock \bibinfo{journal}{BMC infectious diseases} \bibinfo{volume}{17}
  (\bibinfo{year}{2017}) \bibinfo{pages}{612}.
\bibitem[{Smith et~al.(2018)Smith, Trauer, Gambhir, Richards, Maude, Keith, and
  Flegg}]{smith2018agent}
\bibinfo{author}{N.~R. Smith}, \bibinfo{author}{J.~M. Trauer},
  \bibinfo{author}{M.~Gambhir}, \bibinfo{author}{J.~S. Richards},
  \bibinfo{author}{R.~J. Maude}, \bibinfo{author}{J.~M. Keith},
  \bibinfo{author}{J.~A. Flegg},
\newblock \bibinfo{title}{Agent-based models of malaria transmission: a
  systematic review},
\newblock \bibinfo{journal}{Malaria journal} \bibinfo{volume}{17}
  (\bibinfo{year}{2018}) \bibinfo{pages}{299}.
\bibitem[{Mandal et~al.(2011)Mandal, Sarkar, and
  Sinha}]{mandal2011mathematical}
\bibinfo{author}{S.~Mandal}, \bibinfo{author}{R.~R. Sarkar},
  \bibinfo{author}{S.~Sinha},
\newblock \bibinfo{title}{Mathematical models of malaria-a review},
\newblock \bibinfo{journal}{Malaria journal} \bibinfo{volume}{10}
  (\bibinfo{year}{2011}) \bibinfo{pages}{1--19}.
\bibitem[{Macdonald et~al.(1957)}]{macdonald1957epidemiology}
\bibinfo{author}{G.~Macdonald}, et~al.,
\newblock \bibinfo{title}{The epidemiology and control of malaria.},
\newblock \bibinfo{journal}{The Epidemiology and Control of Malaria.}
  (\bibinfo{year}{1957}).
\bibitem[{Aron(1988)}]{aron1988mathematical}
\bibinfo{author}{J.~L. Aron},
\newblock \bibinfo{title}{Mathematical modelling of immunity to malaria},
\newblock \bibinfo{journal}{Mathematical Biosciences} \bibinfo{volume}{90}
  (\bibinfo{year}{1988}) \bibinfo{pages}{385--396}.
\bibitem[{Aron and May(1982)}]{aron1982population}
\bibinfo{author}{J.~L. Aron}, \bibinfo{author}{R.~M. May},
\newblock \bibinfo{title}{The population dynamics of malaria},
\newblock in: \bibinfo{booktitle}{The population dynamics of infectious
  diseases: theory and applications}, \bibinfo{publisher}{Springer},
  \bibinfo{year}{1982}, pp. \bibinfo{pages}{139--179}.
\bibitem[{Dietz et~al.(1974)Dietz, Molineaux, and Thomas}]{dietz1974malaria}
\bibinfo{author}{K.~Dietz}, \bibinfo{author}{L.~Molineaux},
  \bibinfo{author}{A.~Thomas},
\newblock \bibinfo{title}{A malaria model tested in the african savannah},
\newblock \bibinfo{journal}{Bulletin of the World Health Organization}
  \bibinfo{volume}{50} (\bibinfo{year}{1974}) \bibinfo{pages}{347}.
\bibitem[{Hethcote(2000)}]{hethcote2000mathematics}
\bibinfo{author}{H.~W. Hethcote},
\newblock \bibinfo{title}{The mathematics of infectious diseases},
\newblock \bibinfo{journal}{SIAM review} \bibinfo{volume}{42}
  (\bibinfo{year}{2000}) \bibinfo{pages}{599--653}.
\bibitem[{Duan et~al.(2015)Duan, Fan, Zhang, Guo, and
  Qiu}]{duan2015mathematical}
\bibinfo{author}{W.~Duan}, \bibinfo{author}{Z.~Fan},
  \bibinfo{author}{P.~Zhang}, \bibinfo{author}{G.~Guo},
  \bibinfo{author}{X.~Qiu},
\newblock \bibinfo{title}{Mathematical and computational approaches to epidemic
  modeling: a comprehensive review},
\newblock \bibinfo{journal}{Frontiers of Computer Science} \bibinfo{volume}{9}
  (\bibinfo{year}{2015}) \bibinfo{pages}{806--826}.
\bibitem[{Wang and Deisboeck(2013)}]{wang2013agent}
\bibinfo{author}{Z.~Wang}, \bibinfo{author}{T.~S. Deisboeck},
\newblock \bibinfo{title}{Agent-based modeling},
\newblock \bibinfo{journal}{Encyclopedia of Systems Biology}
  (\bibinfo{year}{2013}) \bibinfo{pages}{13--13}.
\bibitem[{Nejat and Damnjanovic(2012)}]{nejat2012agent}
\bibinfo{author}{A.~Nejat}, \bibinfo{author}{I.~Damnjanovic},
\newblock \bibinfo{title}{Agent-based modeling of behavioral housing recovery
  following disasters},
\newblock \bibinfo{journal}{Computer-Aided Civil and Infrastructure
  Engineering} \bibinfo{volume}{27} (\bibinfo{year}{2012})
  \bibinfo{pages}{748--763}.
\bibitem[{Jindal and Rao(2017)}]{jindal2017agent}
\bibinfo{author}{A.~Jindal}, \bibinfo{author}{S.~Rao},
\newblock \bibinfo{title}{Agent-based modeling and simulation of mosquito-borne
  disease transmission},
\newblock in: \bibinfo{booktitle}{Proceedings of the 16th Conference on
  Autonomous Agents and MultiAgent Systems},
  \bibinfo{organization}{International Foundation for Autonomous Agents and
  Multiagent Systems}, \bibinfo{year}{2017}, pp. \bibinfo{pages}{426--435}.
\bibitem[{Mordecai et~al.(2013)Mordecai, Paaijmans, Johnson, Balzer, Ben-Horin,
  de~Moor, McNally, Pawar, Ryan, Smith et~al.}]{mordecai2013optimal}
\bibinfo{author}{E.~A. Mordecai}, \bibinfo{author}{K.~P. Paaijmans},
  \bibinfo{author}{L.~R. Johnson}, \bibinfo{author}{C.~Balzer},
  \bibinfo{author}{T.~Ben-Horin}, \bibinfo{author}{E.~de~Moor},
  \bibinfo{author}{A.~McNally}, \bibinfo{author}{S.~Pawar},
  \bibinfo{author}{S.~J. Ryan}, \bibinfo{author}{T.~C. Smith}, et~al.,
\newblock \bibinfo{title}{Optimal temperature for malaria transmission is
  dramatically lower than previously predicted},
\newblock \bibinfo{journal}{Ecology letters} \bibinfo{volume}{16}
  (\bibinfo{year}{2013}) \bibinfo{pages}{22--30}.
\bibitem[{gad(2019)}]{gadm2019}
\bibinfo{title}{Gadm.},
\newblock
  \bibinfo{journal}{Retrieved~10~October~2019,~from:~\url{https://gadm.org/data.html}}
   (\bibinfo{year}{2019}).
\bibitem[{Gething et~al.(2011)Gething, Patil, Smith, Guerra, Elyazar, Johnston,
  Tatem, and Hay}]{gething2011new}
\bibinfo{author}{P.~W. Gething}, \bibinfo{author}{A.~P. Patil},
  \bibinfo{author}{D.~L. Smith}, \bibinfo{author}{C.~A. Guerra},
  \bibinfo{author}{I.~R. Elyazar}, \bibinfo{author}{G.~L. Johnston},
  \bibinfo{author}{A.~J. Tatem}, \bibinfo{author}{S.~I. Hay},
\newblock \bibinfo{title}{A new world malaria map: Plasmodium falciparum
  endemicity in 2010},
\newblock \bibinfo{journal}{Malaria journal} \bibinfo{volume}{10}
  (\bibinfo{year}{2011}) \bibinfo{pages}{378}.
\bibitem[{Dev et~al.(2015)Dev, Adak, Singh, Nanda, and Baidya}]{dev2015malaria}
\bibinfo{author}{V.~Dev}, \bibinfo{author}{T.~Adak}, \bibinfo{author}{O.~P.
  Singh}, \bibinfo{author}{N.~Nanda}, \bibinfo{author}{B.~K. Baidya},
\newblock \bibinfo{title}{Malaria transmission in tripura: Disease distribution
  \& determinants},
\newblock \bibinfo{journal}{The Indian journal of medical research}
  \bibinfo{volume}{142} (\bibinfo{year}{2015}) \bibinfo{pages}{S12}.
\bibitem[{Blumberg and Frean(2007)}]{blumberg2007malaria}
\bibinfo{author}{L.~Blumberg}, \bibinfo{author}{J.~Frean},
\newblock \bibinfo{title}{Malaria control in south africa-challenges and
  successes},
\newblock \bibinfo{journal}{South African Medical Journal} \bibinfo{volume}{97}
  (\bibinfo{year}{2007}) \bibinfo{pages}{1193--1197}.
\bibitem[{Craig et~al.(1999)Craig, Snow, and le~Sueur}]{craig1999climate}
\bibinfo{author}{M.~H. Craig}, \bibinfo{author}{R.~Snow},
  \bibinfo{author}{D.~le~Sueur},
\newblock \bibinfo{title}{A climate-based distribution model of malaria
  transmission in sub-saharan africa},
\newblock \bibinfo{journal}{Parasitology today} \bibinfo{volume}{15}
  (\bibinfo{year}{1999}) \bibinfo{pages}{105--111}.
\bibitem[{Dawaki et~al.(2016)Dawaki, Al-Mekhlafi, Ithoi, Ibrahim, Atroosh,
  Abdulsalam, Sady, Elyana, Adamu, Yelwa et~al.}]{dawaki2016nigeria}
\bibinfo{author}{S.~Dawaki}, \bibinfo{author}{H.~M. Al-Mekhlafi},
  \bibinfo{author}{I.~Ithoi}, \bibinfo{author}{J.~Ibrahim},
  \bibinfo{author}{W.~M. Atroosh}, \bibinfo{author}{A.~M. Abdulsalam},
  \bibinfo{author}{H.~Sady}, \bibinfo{author}{F.~N. Elyana},
  \bibinfo{author}{A.~U. Adamu}, \bibinfo{author}{S.~I. Yelwa}, et~al.,
\newblock \bibinfo{title}{Is nigeria winning the battle against malaria?
  prevalence, risk factors and kap assessment among hausa communities in kano
  state},
\newblock \bibinfo{journal}{Malaria journal} \bibinfo{volume}{15}
  (\bibinfo{year}{2016}) \bibinfo{pages}{351}.
\bibitem[{com(2019)}]{communicable2019diseases}
\bibinfo{title}{Communicable~diseases~communique},
\newblock
  \bibinfo{journal}{Retrieved~from:~\url{http://www.nicd.ac.za/wp-content/uploads/2017/06/Malaria-in-SA.pdf}}
   (\bibinfo{year}{2019}).
\bibitem[{Adeyemo et~al.(2013)Adeyemo, Makinde, Chukwuka, and
  Oyana}]{adeyemo2013incidence}
\bibinfo{author}{F.~Adeyemo}, \bibinfo{author}{O.~Makinde},
  \bibinfo{author}{L.~Chukwuka}, \bibinfo{author}{E.~Oyana},
\newblock \bibinfo{title}{Incidence of malaria infection among the
  undergraduates of university of benin (uniben), benin city, nigeria},
\newblock \bibinfo{journal}{The Internet Journal of Tropical Medicine}
  \bibinfo{volume}{9} (\bibinfo{year}{2013}) \bibinfo{pages}{1--8}.
\bibitem[{cli(2019)}]{climate2019data}
\bibinfo{title}{Climate~data~for~cities~worldwide-climate-data.org.},
\newblock \bibinfo{journal}{Retrieved~from:~\url{https://en.climate-data.org/}}
   (\bibinfo{year}{2019}).
\bibitem[{Ohm et~al.(2018)Ohm, Baldini, Barreaux, Lefevre, Lynch, Suh,
  Whitehead, and Thomas}]{ohm2018rethinking}
\bibinfo{author}{J.~R. Ohm}, \bibinfo{author}{F.~Baldini},
  \bibinfo{author}{P.~Barreaux}, \bibinfo{author}{T.~Lefevre},
  \bibinfo{author}{P.~A. Lynch}, \bibinfo{author}{E.~Suh},
  \bibinfo{author}{S.~A. Whitehead}, \bibinfo{author}{M.~B. Thomas},
\newblock \bibinfo{title}{Rethinking the extrinsic incubation period of malaria
  parasites},
\newblock \bibinfo{journal}{Parasites \& vectors} \bibinfo{volume}{11}
  (\bibinfo{year}{2018}) \bibinfo{pages}{178}.
\bibitem[{Nanvyat et~al.(2018)Nanvyat, Mulambalah, Barshep, Ajiji, Dakul, and
  Tsingalia}]{nanvyat2018malaria}
\bibinfo{author}{N.~Nanvyat}, \bibinfo{author}{C.~Mulambalah},
  \bibinfo{author}{Y.~Barshep}, \bibinfo{author}{J.~Ajiji},
  \bibinfo{author}{D.~Dakul}, \bibinfo{author}{H.~Tsingalia},
\newblock \bibinfo{title}{Malaria transmission trends and its lagged
  association with climatic factors in the highlands of plateau state,
  nigeria},
\newblock \bibinfo{journal}{Tropical parasitology} \bibinfo{volume}{8}
  (\bibinfo{year}{2018}) \bibinfo{pages}{18}.
\bibitem[{Box(1976)}]{box1976science}
\bibinfo{author}{G.~E. Box},
\newblock \bibinfo{title}{Science and statistics},
\newblock \bibinfo{journal}{Journal of the American Statistical Association}
  \bibinfo{volume}{71} (\bibinfo{year}{1976}) \bibinfo{pages}{791--799}.
\bibitem[{Kim(2015)}]{kim2015t}
\bibinfo{author}{T.~K. Kim},
\newblock \bibinfo{title}{T test as a parametric statistic},
\newblock \bibinfo{journal}{Korean journal of anesthesiology}
  \bibinfo{volume}{68} (\bibinfo{year}{2015}) \bibinfo{pages}{540}.
\bibitem[{Modu et~al.(2016)Modu, Asyhari, and Peng}]{modu2016data}
\bibinfo{author}{B.~Modu}, \bibinfo{author}{A.~T. Asyhari},
  \bibinfo{author}{Y.~Peng},
\newblock \bibinfo{title}{Data analytics of climatic factor influence on the
  impact of malaria incidence},
\newblock in: \bibinfo{booktitle}{2016 IEEE Symposium Series on Computational
  Intelligence (SSCI)}, \bibinfo{organization}{IEEE}, \bibinfo{year}{2016}, pp.
  \bibinfo{pages}{1--8}.
\bibitem[{Parham et~al.(2012)Parham, Pople, Christiansen-Jucht, Lindsay,
  Hinsley, and Michael}]{parham2012modeling}
\bibinfo{author}{P.~E. Parham}, \bibinfo{author}{D.~Pople},
  \bibinfo{author}{C.~Christiansen-Jucht}, \bibinfo{author}{S.~Lindsay},
  \bibinfo{author}{W.~Hinsley}, \bibinfo{author}{E.~Michael},
\newblock \bibinfo{title}{Modeling the role of environmental variables on the
  population dynamics of the malaria vector anopheles gambiae sensu stricto},
\newblock \bibinfo{journal}{Malaria Journal} \bibinfo{volume}{11}
  (\bibinfo{year}{2012}) \bibinfo{pages}{271}.
\bibitem[{cen(2019)}]{census2019data}
\bibinfo{title}{North~tripura~district~population~census~2011-2019,~tripura~
  literacy~sex~ratio~and~density.},
\newblock
  \bibinfo{journal}{[online]~Available~at:~\url{https://census2011.co.in/census/district/460-north-tripura.html}~[Accessed
  17 Jul. 2019].}  (\bibinfo{year}{2019}).
\bibitem[{onl(2019{\natexlab{a}})}]{online2019tripura}
\bibinfo{title}{Outline~of~tripura},
\newblock
  \bibinfo{journal}{Retrieved~17~July~2019,~from:~\url{https://en.wikipedia.org/wiki/Outline-of-Tripura}}
   (\bibinfo{year}{2019}{\natexlab{a}}).
\bibitem[{onl(2019{\natexlab{b}})}]{online2019limpopo}
\bibinfo{title}{Limpopo~population},
\newblock
  \bibinfo{journal}{Retrieved~17~July~2019,~from:~\url{http://population.city/south-africa/adm/limpopo/}}
   (\bibinfo{year}{2019}{\natexlab{b}}).
\bibitem[{onl(2019{\natexlab{c}})}]{online2019gauteng}
\bibinfo{title}{Gauteng},
\newblock
  \bibinfo{journal}{Retrieved~17~July~2019,~from:~\url{https://en.wikipedia.org/wiki/Gauteng}}
   (\bibinfo{year}{2019}{\natexlab{c}}).
\bibitem[{Earth(2019)}]{online2019population}
\bibinfo{author}{N.~P. C. P. N. F. p. L. .~H.~P. Earth, P.},
\newblock \bibinfo{title}{Population~of~benin~city~in~2018-statistics},
\newblock
  \bibinfo{journal}{Retrieved~17~July~2019,~from:~\url{https://all-populations.com/en/ng/population-of-benin-city.html}}
   (\bibinfo{year}{2019}).
\bibitem[{onl(2019)}]{online2019benin}
\bibinfo{title}{Life~expectancy~in~benin},
\newblock
  \bibinfo{journal}{Retrieved~17~July~2019,~from:~\url{https://www.worldlifeexpectancy.com/benin-life-expectancy}}
   (\bibinfo{year}{2019}).
\bibitem[{Lou and Zhao(2010)}]{lou2010climate}
\bibinfo{author}{Y.~Lou}, \bibinfo{author}{X.-Q. Zhao},
\newblock \bibinfo{title}{A climate-based malaria transmission model with
  structured vector population},
\newblock \bibinfo{journal}{SIAM Journal on Applied Mathematics}
  \bibinfo{volume}{70} (\bibinfo{year}{2010}) \bibinfo{pages}{2023--2044}.
\bibitem[{Niger and Gumel(2008)}]{niger2008mathematical}
\bibinfo{author}{A.~M. Niger}, \bibinfo{author}{A.~B. Gumel},
\newblock \bibinfo{title}{Mathematical analysis of the role of repeated
  exposure on malaria transmission dynamics},
\newblock \bibinfo{journal}{Differential Equations and Dynamical Systems}
  \bibinfo{volume}{16} (\bibinfo{year}{2008}) \bibinfo{pages}{251--287}.
\bibitem[{Chitnis et~al.(2006)Chitnis, Cushing, and
  Hyman}]{chitnis2006bifurcation}
\bibinfo{author}{N.~Chitnis}, \bibinfo{author}{J.~M. Cushing},
  \bibinfo{author}{J.~Hyman},
\newblock \bibinfo{title}{Bifurcation analysis of a mathematical model for
  malaria transmission},
\newblock \bibinfo{journal}{SIAM Journal on Applied Mathematics}
  \bibinfo{volume}{67} (\bibinfo{year}{2006}) \bibinfo{pages}{24--45}.
\bibitem[{ven(2019)}]{vensim2019software}
\bibinfo{title}{Vensim~software},
\newblock
  \bibinfo{journal}{Retrieved~from:~\url{https://vensim.com/vensim-software/}}
  (\bibinfo{year}{2019}).
\bibitem[{Wilensky and Rand(2015)}]{wilensky2015introduction}
\bibinfo{author}{U.~Wilensky}, \bibinfo{author}{W.~Rand}, \bibinfo{title}{An
  introduction to agent-based modeling: modeling natural, social, and
  engineered complex systems with NetLogo}, \bibinfo{publisher}{MIT Press},
  \bibinfo{year}{2015}.
\bibitem[{Abar et~al.(2017)Abar, Theodoropoulos, Lemarinier, and
  O’Hare}]{abar2017agent}
\bibinfo{author}{S.~Abar}, \bibinfo{author}{G.~K. Theodoropoulos},
  \bibinfo{author}{P.~Lemarinier}, \bibinfo{author}{G.~M. O’Hare},
\newblock \bibinfo{title}{Agent based modelling and simulation tools: A review
  of the state-of-art software},
\newblock \bibinfo{journal}{Computer Science Review} \bibinfo{volume}{24}
  (\bibinfo{year}{2017}) \bibinfo{pages}{13--33}.
\bibitem[{Cardinot et~al.(2019)Cardinot, O’Riordan, Griffith, and
  Perc}]{cardinot2019evoplex}
\bibinfo{author}{M.~Cardinot}, \bibinfo{author}{C.~O’Riordan},
  \bibinfo{author}{J.~Griffith}, \bibinfo{author}{M.~Perc},
\newblock \bibinfo{title}{Evoplex: A platform for agent-based modeling on
  networks},
\newblock \bibinfo{journal}{SoftwareX} \bibinfo{volume}{9}
  (\bibinfo{year}{2019}) \bibinfo{pages}{199--204}.
\bibitem[{Pollitt et~al.(2015)Pollitt, Bram, Blanford, Jones, and
  Read}]{pollitt2015existing}
\bibinfo{author}{L.~C. Pollitt}, \bibinfo{author}{J.~T. Bram},
  \bibinfo{author}{S.~Blanford}, \bibinfo{author}{M.~J. Jones},
  \bibinfo{author}{A.~F. Read},
\newblock \bibinfo{title}{Existing infection facilitates establishment and
  density of malaria parasites in their mosquito vector},
\newblock \bibinfo{journal}{PLoS pathogens} \bibinfo{volume}{11}
  (\bibinfo{year}{2015}) \bibinfo{pages}{e1005003}.
\bibitem[{Bakare(2015)}]{bakare2015optimal}
\bibinfo{author}{E.~Bakare},
\newblock \bibinfo{title}{On the optimal control of vaccination and treatments
  for an sir-epidemic model with infected immigrants},
\newblock \bibinfo{journal}{Applied \& Computational Mathematics}
  \bibinfo{volume}{4} (\bibinfo{year}{2015}) \bibinfo{pages}{01--09}.
\bibitem[{Changyong et~al.(2014)Changyong, Hongyue, Naiji, Tian, Hua, Ying
  et~al.}]{changyong2014log}
\bibinfo{author}{F.~Changyong}, \bibinfo{author}{W.~Hongyue},
  \bibinfo{author}{L.~Naiji}, \bibinfo{author}{C.~Tian},
  \bibinfo{author}{H.~Hua}, \bibinfo{author}{L.~Ying}, et~al.,
\newblock \bibinfo{title}{Log-transformation and its implications for data
  analysis},
\newblock \bibinfo{journal}{Shanghai archives of psychiatry}
  \bibinfo{volume}{26} (\bibinfo{year}{2014}) \bibinfo{pages}{105}.

\end{thebibliography}

\end{document}